\documentclass[preprint,aps,superscriptaddress,nofootinbib,eqsecnum,%
tightenlines]{revtex4}
\usepackage{amsmath,bm,epsf,mathrsfs}
\newcommand{\<}{\langle}
\renewcommand{\>}{\rangle}
\renewcommand{\d}{\partial}

\newcommand{\x}{\bm{x}}
\newcommand{\y}{\bm{y}}
\newcommand{\q}{\bm{q}}

\newcommand\chiB{\chi_{\raisebox{-0.13em}{\scriptsize B}}}
\newcommand\chiBinv{\chi_{\textrm{B}}^{-1}}
\newcommand\mq{m_{\textrm{q}}}
\newcommand{\bfnab}{\mbox{\boldmath$\nabla$}}
\newcommand\B{{\mathrm{B}}}
\newcommand\I{{\mathrm{I}}}

\begin{document}

\preprint{INT-PUB 04-01}
\title{Dynamic universality class of the QCD critical point}

\affiliation{Institute for Nuclear Theory,
University of Washington, Seattle, Washington 98195-1550}
\affiliation{Department of Physics, University of Illinois, Chicago, 
Illinois 60607-7059}
\affiliation{RIKEN-BNL Research Center, Brookhaven National Laboratory,
Upton, New York 11973}

\author{D.~T.~Son}
\email{son@phys.washington.edu}
\affiliation{Institute for Nuclear Theory,
University of Washington, Seattle, Washington 98195-1550}
\author{M.~A.~Stephanov}
\email{misha@uic.edu}
\affiliation{Department of Physics, University of Illinois, Chicago, 
Illinois 60607-7059}
\affiliation{RIKEN-BNL Research Center, Brookhaven National Laboratory,
Upton, New York 11973}
\date{January 2004} 

\begin{abstract}
We show that the dynamic universality class of the QCD critical point
is that of model H and discuss the dynamic critical exponents.  We
show that the baryon diffusion rate vanishes at the critical point.
The dynamic critical index $z$ is close to 3.
\end{abstract}
\maketitle

\section{Introduction}

The phase diagram of QCD is a focus of many theoretical
investigations~\cite{RajagopalWilczek}. Heavy-ion collision
experiments can probe a certain part of this phase diagram. The temperature
and the baryon chemical potential at freezeout are determined, using
statistical model fits to particle yields and spectra, to span the
domain $T\sim
100-180$~MeV and $\mu_{\rm B}=50-600$~MeV \cite{BMStachel}.  This is the
region where the critical point of QCD, as strongly suggested by
model calculations~\cite{models} and lattice Monte Carlo
results~\cite{lattice}, is located. The critical point is an end point
of a first order phase transition line separating, in the chiral
limit, the chirally broken and chirally symmetric phases.  The precise
location of the critical point is still unknown.

Experimental signatures, based on the singular behavior of
thermodynamic functions near the critical point, were suggested in
Refs.~\cite{SRS,SRS2}. The characteristic feature of all such
signatures is the non-monotonic dependence on the value of experimentally
controlled parameter, such as $\sqrt s$, as the critical point is
approached and passed.  An attempt to estimate the effects of
critical dynamics on experimental observables was made in
Ref.~\cite{SRS2}, where it has been shown that the effects can easily
exceed the experimental background by orders of magnitude. The major
limiting factor is the finite size and time effects, which round up
the critical singularity in all observables.

Near criticality, the crucial quantity is the value of the largest
correlation length $\xi$. For example, the singular contributions to
event-by-event fluctuation observables studied in Ref.~\cite{SRS2} are
proportional to $\xi^2$.  The divergence of $\xi$ is limited by two
effects: i) the finite system size, ${\cal O}(10\ {\rm fm})$ for heavy-ion
collisions, and ii) the finite evolution time.  As pointed out in
Refs.~\cite{SRS2,Berdnikov}, the second effect is more important.
Indeed, the time during which the the correlation length reaches its
equilibrium value diverges as $\tau\sim \xi^z$, which defines the
dynamic scaling exponent $z$.  The finite evolution time limits the
correlation length to be $\xi<{\rm (time)}^{1/z}$.  Since $z>1$ and
typical evolution times of the heavy ion collisions are of the same
order as the spatial size, the time limit is more stringent then
spatial size limitation $\xi<{\rm (size)}$. The value of $z$ depends on the
dynamic universality class~\cite{HH} of the critical point.

The purpose of this paper is to determine this universality class.
While the {\em static} universality class is beyond doubt that of the 3d 
Ising model,
the question about the dynamic universality class has not been 
satisfactorily answered in the literature.
Some arguments were given in Ref.~\cite{SRS2} that the universality
class could be that of model H in Hohenberg and Halperin's
classification~\cite{HH}, -- a model of the liquid-gas phase
transition.  The question was revisited in Ref.~\cite{Berdnikov},
where it was argued that the universality class is that of model C.
The problem is not trivial because of an additional
mode compared to the ordinary liquid-gas phase transition --- the
QCD chiral condensate $\bar qq$.

We shall see that mixing between the chiral condensate $\bar qq$ and
the baryon density $n$ leaves out only one truly hydrodynamic mode --- a
linear combination of the two. The hydrodynamic mode is
conserved and couples to the energy-momentum density.  Consequently,
the relevant hydrodynamic modes near the critical point are the same
as in the model H.  Thus we predict $z\approx 3$.

Our results are in qualitative agreement with the arguments of Fujii
based on a recent study \cite{Fujii}. The study \cite{Fujii} employed
model calculations, and did not, in particular, take into account
the coupling to energy-momentum density. Our approach is
model-independent and is based on the straightforward application of
the hydrodynamic theory, reviewed, e.g., in Ref.~\cite{LL5}.

The paper is organized as follows. In Sec.~\ref{sec:modes} we identify
the set of hydrodynamic modes near the critical point.
In Sec.~\ref{sec:no-j} we discuss a
simplified problem, where we forcibly neglect the energy-momentum
density.  The purpose is to demonstrate that only one linear
combination of the chiral order parameter $\langle\bar qq\rangle$ and
the baryon number density $n$ is a true hydrodynamic mode.  To keep 
the discussion simple, we
shall work at the mean field level, sufficient to identify the
correct universality class. We refer to the literature for
going beyond mean field.  In
Sec.~\ref{sec:j} we introduce the energy-momentum density, restoring
the full set of hydrodynamic variables near the QCD critical point.
We show that the relevant modes in the long-length, long-time limit
are the same as in the universality class of the liquid-gas phase
transition. We conclude with Sec.~\ref{sec:concl}.  In 
Appendix~\ref{app:isospin} we
show that the inclusion of the isospin density does not affect the
critical behavior.

\section{Hydrodynamic modes near the QCD critical point}
\label{sec:modes}

Our analysis is based on the premise that at sufficiently large
distance and time scales, the dynamics of a finite-temperature system
is described by a hydrodynamic theory. It is a theory which contains
only the degrees of freedom varying slowly with time.  Although the
validity of hydrodynamics is amply supported by experiments, a generic
theoretical derivation, or proof, of the validity of such a
description is a challenging problem. Nevertheless, the validity
of hydrodynamic description has been verified in certain theories:
scalar field theories~\cite{Jeon} and strongly coupled gauge theories
with gravity duals~\cite{gravity-dual}.  We shall therefore assume
that the dynamics of the finite-temperature QCD plasma can be
described by a set of hydrodynamic equations.

Generally, the full set of hydrodynamic modes include (i) densities of
conserved charges, which relax via diffusion; (ii) phases of
symmetry-breaking condensates; (iii) Abelian gauge fields of unbroken
U(1) gauge symmetries.  Near second-order phase transitions, (iv) the
full order parameter (not only its phases, but also the magnitude)
relax slowly, and therefore should be included in hydrodynamics.
Thus, near the critical point of QCD on the $(T,\mu)$ plane, the modes
potentially important for hydrodynamics are given by the fluctuations of:
\begin{itemize}
\item The conserved energy and momentum densities: 
$\varepsilon\equiv T^{00}-\<T^{00}\>$, and $\pi^i\equiv T^{0i}$;
\item The conserved baryon number density, $n\equiv \bar
  q\gamma^0 q -\<\bar q\gamma^0q\>$;
\item The chiral condensate $\sigma\equiv 
  \bar qq - \<\bar qq\>$.
\end{itemize}

The finiteness of the pion mass prevents pions from being a
low-frequency hydrodynamic mode.  This should be contrasted with the
chiral limit, where the massless pions do become hydrodynamic modes,
leading to a different universality class (both static and
dynamic)~\cite{RW-dyn}.  

In QCD, the complete set of conserved charge densities
also includes isospin density (neglecting small isospin-breaking
effects).  However, the coupling of the isospin to other modes turns
out to be irrelevant in the renormalization group sense (see 
Appendix~\ref{app:isospin}).  Therefore
we do not need to include the isospin density in our treatment.

\section{Without energy-momentum tensor}
\label{sec:no-j}

\subsection{Statics}

We start by discussing the statics.  The static correlation functions
can be found from a Ginzburg-Landau functional
\begin{equation}\label{GL}
   F[\sigma,n] = \int\!d\x\, \Bigl[\frac a2(\d_i \sigma)^2+
  b\d_i\sigma\d_i n + \frac c2(\d_i n)^2 
  + V(\sigma, n) \Bigr],
\end{equation}
where
\begin{equation}
  V(\sigma,n) = \frac A2\sigma^2 + B\sigma n + \frac C2 n^2
  + \mbox{terms of higher orders}.
\end{equation}
The equilibrium distribution of the system is governed by $e^{-\beta
F}$.  The mixing between $\sigma$ and $n$ is not forbidden at nonzero
baryon chemical potential and quark masses, and has been taken into
account.

As we approach the phase transition, all the parameters of the
Ginzburg-Landau functional ($a$, $b$, $c$, $A$, $B$, $C$) remain
finite, but the quadratic form in $V$ becomes degenerate, $AC=B^2$.
At the phase transition, the potential has zero curvature along one
direction in the $(\sigma, n)$ plane: 
\begin{equation}\label{flat}
 \mbox{flat direction:}\qquad \frac{\sigma}{n} = -\frac BA=-\frac CB\,.
\end{equation}

The response of the system to static external perturbation can be
found by including source terms into the Ginzburg-Landau free energy.
For example, changing the chemical potential from $\mu$ to
$\mu+\delta\mu$ and/or the quark mass from $\mq$ to $\mq+\delta\mq$
induces the following change in the average variables:
\begin{subequations}
 \begin{eqnarray}
  \sigma = &\Delta^{-1}(\,- B\, \delta\mu\ - &C \, \delta \mq\,)\,,
\\
   n = &\Delta^{-1}(\, +A\, \delta\mu\ + & B\, \delta \mq\,)\,.
\end{eqnarray} 
\end{subequations}
where 
\begin{equation}
   \Delta = AC - B^2.
\end{equation}
At the critical point $\Delta=0$ and the responses are singular.
In particular, the baryon susceptibility $\chiB\equiv dn/d\mu$ 
diverges at the phase
transition.  Near the phase transition, both changes induced by
$\delta\mu$ and $\delta m_q$ occur along the
flat direction  (\ref{flat}) of the potential.

From the distribution $e^{-\beta F}$ one can compute the fluctuations
of $\sigma$ and $n$.  In the limit $\q\to0$ the quadratic fluctuations
are
\begin{equation}\label{eqtcorr}
  \<\sigma^2_{\bm q\to0}\> = \frac{TC}\Delta\,, \qquad
  \<n^2_{\bm q\to0}\> = T\chi_B = \frac{TA}\Delta\,.
\end{equation}

Finally, by considering the response to nonuniform perturbations, $\bm
q\ne0$, one can determine the correlation length $\xi$.  
Near the critical point
it diverges as
\begin{equation}
  \xi \sim {\Delta^{-1/2}}.
\end{equation}

\subsection{Hydrodynamic equations}

At the linear order, the hydrodynamic equations can be written by
using the standard rules reviewed in, e.g., Ref.~\cite{LL5} (see also
Ref.~\cite{pion-propagation}).  We
define the variables conjugate to $\sigma$ and $n$:
\begin{subequations}
\begin{eqnarray}
  X_\sigma &=& \frac{\delta F}{\delta \sigma} =
     (A-a\nabla^2)\sigma + (B- b\nabla^2) n\,,\\
  X_n &=& \frac{\delta F}{\delta n} = 
     (B-b\nabla^2)\sigma + (C- c\nabla^2) n\,.
\end{eqnarray}
\end{subequations}
The linearized equations for $\sigma$ and $n$ are then
\begin{subequations}
\begin{eqnarray}
  \dot\sigma(\q) &=& -\gamma_{\sigma\sigma}(\q) X_\sigma(\q) 
                    - \gamma_{\sigma n}(\q) X_n(\q) + \xi_\sigma(\q)\,, \\
  \dot n(\q) &=& -\gamma_{n\sigma}(\q) X_\sigma(\q) - \gamma_{nn}(\q)X_n(\q)
                +\xi_n(\q)\,.
\end{eqnarray}
\end{subequations}
We have written the equations for each spatial momentum $\q$.
Onsager's principle forces $\gamma_{\sigma n}=\gamma_{n\sigma}$.  The
noise correlators are
\begin{equation}
  \<\xi_i(t,\q) \xi_j(t'\!,\q')\> = 2T \gamma_{ij}(\q) 
    (2\pi)^3\delta(\q-\q') \delta(t-t'), \qquad i, j=\sigma, n,
\end{equation}
and are such so that the equilibrium distribution is given by
$e^{-\beta F}$.

In the limit of small momenta $\q\to0$ relevant to hydrodynamics, one
can expand $\gamma_{ij}(\q)$ in powers of $q^2$ and keep only the
leading terms.  Due to the conservation of baryon charge,
$\gamma_{n\sigma}$ and $\gamma_{nn}$ vanish in the limit $\q=0$ and
their expansions start at the order $q^2$.  There is no such
constraint on $\gamma_{\sigma\sigma}$, so the expansion for this
coefficient starts at the $q^0$ order.  Introducing the notations
\begin{subequations}
\begin{eqnarray}
  \gamma_{\sigma\sigma}(\q) &=& \Gamma + O(q^2)\,,\\
  \gamma_{\sigma n}(\q) &=& \tilde\lambda q^2 + O(q^4)\,,\\
  \gamma_{nn}(\q) &=& \lambda q^2 + O(q^4)\,,
\end{eqnarray}
\end{subequations}
the hydrodynamic equations become
\begin{subequations}
\begin{eqnarray}
  \dot\sigma &=& -\Gamma \frac{\delta F}{\delta\sigma} 
   +\tilde\lambda\bfnab^2\frac{\delta F}{\delta n}+ \xi_\sigma\,,\\
  \dot n &=& \tilde\lambda\bfnab^2\frac{\delta F}{\delta\sigma}
   + \lambda\bfnab^2\frac{\delta F}{\delta n} + \xi_n\,.
\end{eqnarray}
\end{subequations}
with the noise correlators
\begin{subequations}
\begin{eqnarray}
  \<\xi_\sigma(x)\xi_\sigma(y)\> &=& 2T\Gamma\delta^4(x-y)\,,\\
  \<\xi_\sigma(x)\xi_n(y)\> &=& -2T\tilde\lambda
      \delta(t-t')\bfnab^2\delta^3(\x-\y)\,,\\
  \<\xi_n(x)\xi_n(y)\> &=& -2T\lambda\delta(t-t')\bfnab^2\delta^3(\x-\y)\,.
\end{eqnarray}
\end{subequations}

\subsection{Modes}

Now let us insert the expression (\ref{GL}) for $F$ into the
hydrodynamic equations and find the dispersion relations for the
normal modes.  To leading order in the limit $\q\to0$ we find
\begin{subequations}
\begin{eqnarray}
  \dot \sigma &=& -\Gamma A\sigma -\Gamma B n\,,\\
  \dot n      &=& (\tilde\lambda A+\lambda B) \bfnab^2\sigma 
      + (\tilde\lambda B+\lambda C)\bfnab^2 n\,.
\end{eqnarray}
\end{subequations}
The eigenfrequencies of the hydrodynamic equations are found by
solving the equation
\begin{equation}
  \det\left|\begin{array}{ccc} \Gamma A-i\omega & & \Gamma B \\ 
  (\tilde\lambda A + \lambda B)q^2 & & 
  (\tilde\lambda B + \lambda C)q^2-i\omega\end{array}\right| = 0\,.
\end{equation}
Near the critical point the two eigenfrequencies are
\begin{subequations}
\begin{eqnarray}
  \omega_1 &=& -i\lambda\frac {\Delta}A \q^2\,.\\
  \omega_2 &=& -i\Gamma A\,.
\end{eqnarray}
\end{subequations}
Therefore, for small $\q$ there are two frequency scales: a small
scale $\propto q^2$ and a larger scale $\propto q^0$.  In the limit
$q\to0$ only $\omega_1$ is truly hydrodynamic.  It is a diffusive mode
with the diffusion constant
\begin{equation}\label{antidiffusion}
  D = \lambda\frac {\Delta}A = \lambda C - \lambda \frac{B^2}A\,.
\end{equation}

The diffusion constant $D$ tends to 0 as the critical point
is approached, $\Delta\to0$.  To understand this fact consider
$\Delta<0$.  In this case the potential $V$ has an unstable direction.
Negative $D$ means that, instead of relaxing, perturbations grow,
leading to spinodal decomposition. The term  $\lambda B^2/A$ in
Eq.~(\ref{antidiffusion}) can be interpreted as ``antidiffusion'' ---
at the critical point it exactly cancels out the normal diffusion term
$\lambda C$.

To interpret the modes corresponding to $\omega_1$ and $\omega_2$, we
find the eigenmodes of the hydrodynamic equations.  Near $q=0$
the two eigenmodes are
\begin{equation}
  \left( \begin{array}{c} -B \\ A\end{array}\right) \qquad \textrm{and}
  \qquad \left( \begin{array}{c} 1 \\ 0\end{array}\right).
\end{equation}
At the critical point, the first mode is the flat direction
(\ref{flat}) of the potential energy $V$.  
Both $\sigma$ and $n$ vary along this direction.  On
the other hand, the mode with frequency $\omega_2$ corresponds to
fluctuations of $\sigma$ alone, unaccompanied by any change of $n$.

The meaning of the two modes is simple. A generic smooth perturbation
involving both $\sigma$ and $n$ will relax over two distinct time
scales. First, at a much shorter time scale, $(\Gamma A)^{-1}$, the
field $\sigma(\x)$ alone adjusts to the values which at each point
$\x$ in space minimize the potential $V(\sigma,n)$ at a given local
value of $n(\x)$: $\sigma=-(B/A)n$. After that, at longer time scales,
the chiral order parameter $\sigma$ can be forgotten. It simply
``traces'' the profile of $n$, which relaxes to $n=0$ much
slower, over the diffusive time scale $(D\q^2)^{-1}$. From the point of view
of slow dynamics at long distances, the addition of the chiral order
parameter $\sigma$ does not change the physics, which is exactly that
of the liquid-gas phase transition, with a conserved density as the
only hydrodynamic mode (besides the energy-momentum).

\subsection{Real-time correlators}
From the stochastic equations one can compute the real-time
correlators of $\sigma$ and $n$. Assuming we are in the regime $D
q^2\ll\Gamma A$, the correlators can be written as
\begin{subequations}
\begin{eqnarray}
  \<\sigma^2_{\omega\q}\> &=& \frac{2T\Gamma}{\omega^2+\Gamma^2A^2}
  + \frac{2TB^2\lambda q^2}{A^2(\omega^2+D^2q^4)}\,,\\
  \<n^2_{\omega\q}\> &=& \frac{2T\lambda q^2}{\omega^2+D^2q^4}\,.
\end{eqnarray}
\end{subequations}
Integrating over $\omega$ we recover the equal-time correlators 
(\ref{eqtcorr}).
Notice that while $\<\sigma^2\>$ obtains contribution from two
characteristic scales $\omega\sim\Gamma A$ and $\omega\sim Dq^2$,
the fluctuations of $n$ are peaked only at the latter scale.

\subsection{Baryon diffusion, susceptibility  and critical indices in $d=3$}

We have seen that, from the point of view of real-time dynamics,
the criticality is manifested by vanishing of the baryon diffusion
rate $D$ given by Eq. (\ref{antidiffusion}). Comparing to
(\ref{eqtcorr}), we find that
\begin{equation}\label{Dchi}
  D=\lambda\chiBinv.
\end{equation}
This relation has a simple physical explanation.
Consider a configuration where $n$ varies
with space, $n=n(\x)$.  This corresponds to a spatially varying
chemical potential $\mu(\x)=\chiBinv n(\x)$. Recall that $\mu$
acts as the time component $A_0$ of an external
gauge field coupled to the baryon current. Thus such a
chemical potential $\mu(\x)$ corresponds to an external electric field ${\bm
E}=-\bfnab\mu$ acting on the baryon charge. The baryon current
induced by this field is ${\bm j}_n=\lambda {\bm E} =
-\lambda\chiBinv\bfnab n$, where $\lambda$ is the baryon conductivity.
Using the definition of diffusion rate $\bm j_n = -D \bfnab n$ we 
obtain (\ref{Dchi}).

At $d=3$ one has to take into account fluctuations which will modify
the numerical values of the critical exponents.  
With the
energy-momentum density fluctuations frozen, as in this section, the
universality class is that of model B in the Hohenberg-Halperin
classification \cite{HH}. It is a model of the uniaxial ferromagnet
--- with a single conserved mode.
In the model B the diffusion constant
is still inversely proportional to the susceptibility, as in Eq.~(\ref{Dchi}):
\begin{equation}\label{D-modelB}
  D \sim \chiBinv \sim \xi^{-2+\eta},
\end{equation}
but the divergence of $\chi_B$ is modified by
$\eta\approx0.04$ --- a critical index of the 3d Ising model,
relative to the mean field result.

The dispersion relation $\omega=-iDq^2$ applies only in the regime
$q\ll\xi^{-1}$.  When $q\gg\xi^{-1}$, we are in the critical regime, and
higher powers of $q$ are non-negligible. In this regime the scaling dictates
dispersion $\omega\sim q^z$.  The two behaviors should match
smoothly at $q\sim\xi^{-1}$.  From Eq.~(\ref{D-modelB}) one then finds
$z=4-\eta$ (in mean-field theory $z=4$).

\section{Coupling to energy-momentum}
\label{sec:j}

So far we have completely neglected the motion of the plasma,
regarding the latter as a static medium.  Now we will allow
the plasma to move --- this will modify the hydrodynamic equations and
change the value of the dynamic critical exponent.

As we have learned in the previous Section, only one combination of
$\sigma$ and $n$ is truly hydrodynamic.  This mode can be thought of
as the baryon charge density $n$, with $\sigma$ simply tracing $n$ and
not independent.  At sufficiently long timescales, 
the hydrodynamic theory contains this mode, the
energy density $\varepsilon$ and the momentum density $\pi^i$. The
linearized theory has four eigenmodes: the baryon diffusion mode, two
 diffusive transverse shear modes and a propagating 
longitudinal sound wave.

At very low momenta, and long time scales, sound wave mode can be
effectively integrated out, since it has linear dispersion $\omega\sim
q$ and its frequency is much higher than the frequencies of the
remaining diffusive modes $\omega\sim q^2$. In other words, the
fluctuations of the energy density and pressure caused by sound excitations
are fast and average out --- there are no sound waves on these long 
diffusive time scales.
The number of remaining hydrodynamic modes is three: 
two transverse components of $\pi^i$ and the baryon
density $n$.\footnote{The physically intuitive argument presented here is 
corroborated by the analysis of linear mixing of four
variables: $\sigma$, $n$, $\epsilon$ and the longitudinal
component of $\pi^i$. As in Section~\ref{sec:no-j},
there is only one diffusive mode, in which $\sigma$ and $\epsilon$
trace the local value of $n$ to minimize the local value of the 
thermodynamic potential (i.e., maximize the pressure).}

The same set of modes describes hydrodynamics of the liquid-gas
phase transition~\cite{SHH-liquid-gas}.  Here we only review the
results of previous studies, referring the reader to the original
literature~\cite{SHH-liquid-gas,HH} for further details.

The liquid-gas phase transition belongs to the dynamic universality
class of model H.  This model describes a system with a conserved order
parameter, conserved and transverse momentum density, 
and nonzero Poisson bracket
between the two.  The same model describes the critical point of
binary fluids.  The hydrodynamic theory contains two kinetic
coefficients: the shear viscosity $\bar\eta$ and the diffusion constant
(in our case, that of the baryon charge $D$).\footnote{We choose the
notation $\bar\eta$ for the shear viscosity to avoid confusion with
the static critical exponent $\eta$.  The constant of heat conductance
$\kappa$ is not an independent constant, but is related to $D$ by
$D\propto\kappa\chiBinv$.}

Both kinetic coefficients are singular (nonanalytic) at
the phase transition:
\begin{subequations}
\begin{eqnarray}
   D =\lambda\chiBinv &\sim& \xi^{x_\lambda}\chiBinv,\\
   \bar\eta &\sim& \xi^{x_\eta}.
\end{eqnarray}
\end{subequations}
The two new dynamical critical exponents satisfy the following relation
\begin{equation}\label{lambda+eta}
  x_\lambda + x_\eta = 4-d-\eta\,.
\end{equation}
The enhancement of $D$ by a power of $\xi$ is due to the contribution
of convection to the baryon conductivity $\lambda$.  Although the
calculation of $x_\lambda$ and $x_\eta$ individually is not simple,
the relation (\ref{lambda+eta}) has a simple physical
explanation~\cite{HH}. Let us apply the field $\bm E=-\bm
\nabla\mu$ introduced already in Section \ref{sec:no-j}. In model B,
without convection, this field would induce baryon current via
diffusion: $\bm j_n = \lambda \bm E$, where $\lambda$ is
finite. However, in this field $\bm E$, the baryon charge carrying
fluid will experience mechanical force equal to $n \bm E$ per unit
volume. The fluid will accelerate to velocity $\bm v$ at which viscous
drag balances the external force:
$  \bm f_{\rm visc}+\bm f_{\rm appl}=0$\,.
For a chunk of fluid of typical linear dimension
$L$ the drag and the applied forces are of order:
\begin{equation}
  \bm f_{\rm visc} \sim - \bar\eta \rho \bm v L^{d-2},
\qquad\mbox{and}\qquad
\bm f_{\rm appl} \sim n \bm E L^d,
\end{equation}
where $\rho$ is the mass density.
The corresponding velocity $\bm v$ diverges with $L$ and so does the induced
baryon current:
\begin{equation}
  \bm j_n = n \bm v \sim \frac{n^2}{\bar\eta\rho}\,L^2\,\bm E\,.
\end{equation} 
The divergence is cut off at the scale $L\sim\xi$ because the baryon number
fluctuation, given by
$\langle n^2 \rangle = T\chiB/ L^d$ for $L\gg\xi$,
is correlated on at most that scale.
Therefore at the critical point we find a singular
contribution to the product of the kinetic coefficients
\begin{equation}
  \lambda\bar\eta\sim
\langle n^2\rangle\xi^2 \sim \chiB\xi^{2-d}\sim \xi^{4-d-\eta},
\end{equation}
 which leads 
to Eq.~(\ref{lambda+eta}).

In the $d=3$ Ising model $\eta$ is very small.  It has been 
found~\cite{HH,SHH-liquid-gas,mode-coupling} that
the divergence of the shear viscosity is very weak, or $x_\eta$ is
numerically small.  According to Eq.~(\ref{lambda+eta}), therefore,
$x_\lambda\approx1$.  Calculations using renormalization group in
$4-\epsilon$ dimensions yields the values~\cite{HH,SHH-liquid-gas}
\begin{subequations}
\begin{eqnarray}
  x_\eta    &=& \frac  1 {19}\epsilon(1+0.238\epsilon+\cdots)
    \approx 0.065\,,\\  
  x_\lambda &=& \frac{18}{19}\epsilon(1-0.033\epsilon+\cdots)
    \approx 0.916\,,
\end{eqnarray}
\end{subequations}
while mode-coupling calculations done directly in $d=3$ and assuming
$\eta=0$ yield~\cite{mode-coupling}
\begin{equation}
  x_\eta = 0.054\,,\qquad x_\lambda = 0.946\,.
\end{equation}

One sees that in model H the diffusion constant still vanishes at the
critical point:
\begin{equation}
  D \sim \xi^{-2+\eta+x_\lambda},
\end{equation}
but the power is closer to 1 rather than to 2 as in model B.  That is
due, as we have mentioned, to fluctuations of the fluid motion.  
Using the same argument as the one
presented at the end of Sec.~\ref{sec:no-j}, we find
\begin{equation}
  z = 4 -\eta - x_\lambda \approx 3\,.
\end{equation}

\section{Conclusion}
\label{sec:concl}

We have seen that the dynamic universality class of the QCD critical
point is that of model H, i.e., the liquid-gas phase transition. The
chiral order parameter, being non-conserved and mixing with the
conserved baryon charge, does not affect the dynamic 
universality class.

Here we would also like to compare our results to the universality
class argument of Ref.~\cite{Berdnikov}.  The argument is based on
three assumptions, spelled in Ref.~\cite{Berdnikov}: i) the
chiral order parameter $\sigma$ is not conserved; ii) there are other
conserved quantities, such as the baryon density $n$; iii) the Poisson
brackets between $\sigma$ and the conserved quantities vanish. While the
first two assumptions are correct, the last assumption is not.  There
is a non-zero Poisson bracket between $\sigma$ (or $n$) and the momentum
density $[\pi_i(\bm x),\sigma(\bm y)]=\sigma(\bm
x)\nabla_i\delta^3(\bm x-\bm y)$.  As we have seen, the coupling to
momentum density plays essential role in determining the dynamic
universality
class. Another, more subtle implicit assumption of
Ref.~\cite{Berdnikov} is that the order parameter $\sigma$ and baryon
density $n$ cannot mix. This assumption leads to model C, where the mixing is
forbidden by a symmetry of the order parameter (it would be
$\sigma\to-\sigma$ in this case). As we have seen in
Section~\ref{sec:no-j}, 
the $\sigma n$
mixing eliminates nonconserved mode from hydrodynamic theory, leading
instead to model B.

The model H value of $z\approx3$ is larger than the model C value
$z\approx2.17$ used in Ref.~\cite{Berdnikov}. As discussed in the
introduction, this means that the effect of the time constraint
on the correlation length $\xi$ is stronger. Thus the
numerical estimate of maximal $\xi$ in Ref.~\cite{Berdnikov}
should be revised downward.
However, we do not expect the main conclusions of
Ref.~\cite{Berdnikov} to change qualitatively.%
\footnote{
Rerunning the codes of Ref.~\cite{Berdnikov} with
the new value of $z$ shows \cite{krishna}
that the required revision is numerically small
(less than 10\%) and is within the many uncertainties of the method detailed in
Ref.~\cite{Berdnikov}. We thank Krishna Rajagopal for sharing this
result with us.}

An interesting step
beyond the relaxation equation for the diverging correlation length in
Ref.~\cite{Berdnikov}, which one can contemplate, having the correct 
hydrodynamic theory in hand, is to consider real-time evolution of the
hydrodynamic modes near criticality, similar to the study of Ref.~\cite{DCC}.

The pions do not enter the hydrodynamic theory as long as the
correlation length $\xi$ is sufficiently large compared to the inverse
of the pion mass $1/m_\pi$ (see Section~\ref{sec:modes}). Although this
condition is always fulfilled sufficiently close to the critical
point, in the realistic case of a heavy ion collision, the maximal
achievable $\xi$ ($2-3$ fm, according to \cite{SRS2,Berdnikov}) is not
significantly larger than $1/m_\pi$. Therefore it might be interesting
to study the effects due to the crossover between the two regimes
$\xi\gg1/m_\pi$ (model H) and $\xi\ll1/m_\pi$ (O(4) antiferromagnet
\cite{RW-dyn}).

It would be interesting to explore the phenomenological consequences
of the vanishing baryon diffusion rate for heavy ion collisions.  Most
likely it will manifest itself in the fluctuations of the baryon
number~\cite{HattaStephanov}.

\begin{acknowledgments}

We thank H. Fujii and K.~Rajagopal for discussions 
and for their comments on the manuscript.
M.A.S. thanks RIKEN
BNL Center and U.S. Department of Energy [DE-AC02-98CH10886] for
providing facilities essential for the completion of this work.
D.T.S. is supported, in part, by DOE grant No.\
DOE-ER-41132 and the Alfred P.\ Sloan Foundation.
M.A.S. is
supported, in part, by DOE grant No.\ DE-FG0201ER41195 
and by the Alfred P.\ Sloan Foundation.

\end{acknowledgments}

\appendix

\section{Isospin density and scaling dimension counting}
\label{app:isospin}

In this appendix we show that the inclusion of the isospin density
$n_\I$ does not affect the critical behavior. For simplicity, let us
neglect the energy-momentum tensor and discuss only the coupled system
of the isospin density and the baryon density.  The coupled dynamics
at the critical point is given by a pair of stochastic equations:
\begin{subequations}
\begin{eqnarray}
  \dot n_\B &=& \lambda_\B \bfnab^2 \Bigl(-c\bfnab^2 n_\B + g_\B n_\B^3
  + \frac{g_{\B\I\I}}2 n_\I^2\Bigr) + \xi_\B \label{nBdot}\,;\\
  \dot n_\I &=& \lambda_\I \bfnab^2 \Bigl({\chi_\I^{-1}}{n_\I} 
   + g_\I n_\I^3 + g_{\B\I\I} n_\B n_\I\Bigr) + \xi_\I\,. \label{nIdot}
\end{eqnarray}
\end{subequations}
Note that the term $\lambda_\B \chi_\B^{-1} \bfnab^2 n_\B$ in
Eq.~(\ref{nBdot}) vanishes at the critical point, because
$\chi_\B\to\infty$.  The isospin susceptibility $\chi_\I$ is finite
\cite{HattaStephanov}. We kept only coupling terms of lowest order
(most relevant in the infrared) consistent with the symmetry
$n_\I\to-n_\I$.  The noises $\xi_\B$ and $\xi_\I$ are auto-correlated
as
\begin{subequations}\label{xiBI}
\begin{eqnarray}
  \<\xi_\B(t,\x)\xi_\B(t',\y)\> &=& 
     -2T\lambda_\B\delta(t-t')\nabla^2\delta^d(\x-\y) \,;\\
  \<\xi_\I(t,\x)\xi_\I(t',\y)\> &=& 
     -2T\lambda_\I\delta(t-t')\nabla^2\delta^d(\x-\y)\,.
\end{eqnarray}
\end{subequations}
Here $d$ is the number of spatial dimensions.

The canonical scaling dimensions involved in renormalization group
transformations are determined as follows.  Comparing linear terms in 
Eq.~(\ref{nBdot}) we
establish that if space has dimension $-1$, $[x]=-1$, then the
dimension of time is $[t]=-4$.  The dimension of noises are then found
from Eqs.~(\ref{xiBI}) to be $[\xi_\B]=[\xi_\I]=d/2+3$.  From
Eq.~(\ref{nBdot}) one then finds $[n_\B]=d/2-1$.  In Eq.~(\ref{nIdot})
we see that the term $\dot n_\I$ can be neglected in the infrared, so
$[n_\I]=d/2+1$. 

Using these scaling dimensions for the fields, we can now
determine the scaling dimensions of the coupling terms.
One finds that $[\lambda_\B g_{\B\I\I}]=-1-d/2$, and
$[\lambda_\I g_{\B\I\I}]=1-d/2$, so the coupling between baryon and
isospin densities is irrelevant for $d>2$.  (In contrast, $[\lambda_\B
g_\B]=4-d$, so the $n_\B^3$ term is relevant for $d<4$.)

Similarly one can show that
isospin density remains decoupled from the critical dynamics when the
energy-momentum tensor is included.
This simple power counting scheme can be formalized by rewriting the
stochastic equations as a path integral.

\end{document}